\documentclass[conference,compsoc]{IEEEtran}
\ifCLASSOPTIONcompsoc
  \usepackage[nocompress]{cite}
\else
  \usepackage{cite}
\fi
\ifCLASSINFOpdf
\else
\fi
\usepackage{color}
\usepackage{listings}
\usepackage{tikz}
\usepackage{tikz-cd}
\usepackage{msc}
\usepackage{float}
\usepackage[ruled,linesnumbered]{algorithm2e}
\usepackage{amsmath}
\usepackage{booktabs}
\usepackage{url}

\definecolor{codegray}{rgb}{0.5,0.5,0.5}
\definecolor{codepurple}{rgb}{0.58,0,0.82}
\definecolor{codegreen}{rgb}{0,0.6,0}
\definecolor{codeblue}{rgb}{0.1,0.1,0.7}
\definecolor{codered}{rgb}{0.7,0.1,0.1}

\lstdefinelanguage{json}{
    keywords={true,false,null},
    keywordstyle=\color{codeblue},
    stringstyle=\color{codered},
    commentstyle=\color{codegray},
    morestring=[b]",
    morecomment=[l]{//},
    morecomment=[s]{/*}{*/},
}

\lstdefinelanguage{yaml}{
    keywords={apiVersion, kind, metadata, name, namespace, annotations,
              spec, volumes, projected, sources, serviceAccountToken, path,
              audience, expirationSeconds, volumeMounts, mountPath, readOnly,
              provider_id, aws, account_id, attribute_condition, attribute_mapping,
              google.subject, attribute.aws_role},
    keywordstyle=\color{codeblue},
    stringstyle=\color{codegreen},
    commentstyle=\color{codegray},
    morestring=[b]',
    morestring=[b]",
    morecomment=[l]{\#},
}

\begin{document}
%
\title{A Multi-Cloud Framework for Zero-Trust Workload Authentication}

\author{\IEEEauthorblockN{Saurabh Deochake\IEEEauthorrefmark{1}, Ryan Murphy\IEEEauthorrefmark{1}, and Jeremiah Gearheart\IEEEauthorrefmark{1}}
\IEEEauthorblockA{\IEEEauthorrefmark{1}SentinelOne Inc.\\
Mountain View, California 94041\\
Email: \{saurabh.deochake, ryan.murphy, jeremiah.gearheart\}@sentinelone.com}}


%


\maketitle

\begin{abstract}
Static, long-lived credentials for workload authentication create untenable security risks that violate Zero-Trust principles. This paper presents a multi-cloud framework using Workload Identity Federation (WIF) and OpenID Connect (OIDC) for secretless authentication. Our approach uses cryptographically-verified, ephemeral tokens, allowing workloads to authenticate without persistent private keys and mitigating credential theft. We validate this framework in an enterprise-scale Kubernetes environment, which significantly reduces the attack surface. The model offers a unified solution to manage workload identities across disparate clouds, enabling future implementation of robust, attribute-based access control.
\end{abstract}


%
\IEEEpeerreviewmaketitle

\section{Introduction}\label{sec:intro}

The rapid adoption of cloud-native and multi-cloud architectures has fundamentally altered how organizations build and deploy software, yet this decentralized paradigm creates a fundamental dissonance with legacy security practices. For decades, workload-to-workload communication has been secured using a static, long-lived model where workloads authenticate to services via private keys or other static credentials. In a modern environment with thousands of dynamic workloads, this approach presents an untenable security risk. A single compromised key can function as a "skeleton key" to vast cloud resources, enabling a clear threat model where an adversary's primary goals are data exfiltration, resource hijacking, and lateral movement between cloud environments \cite{Grobauervuln}. For this paper, we assume an attacker capable of compromising an individual workload to steal a credential, or of finding a key accidentally committed to a source code repository. Furthermore, the operational burden of managing, rotating, and auditing these credentials at scale is immense, often leading to insecure shortcuts and over-privileged access that directly violates the core tenets of a Zero-Trust security posture \cite{NIST-ZTA}.

Addressing this critical and scalable challenge requires a new authentication paradigm aligned with the principles of Zero-Trust. Predicated on the philosophy of "never trust, always verify," Zero-Trust demands that every access request be individually authenticated and authorized with the least possible privilege \cite{syedzta}. This paper presents a practical, multi-cloud framework for implementing Zero-Trust workload authentication, designed to fundamentally replace the static credential model \cite{Deochake2022iam} \cite{deochake2022bigbirdbigdatastorage} with a dynamic, cryptographically-verified alternative. Our solution leverages established open standards such as Workload Identity Federation (WIF) \cite{gcp_wif} and OpenID Connect (OIDC) \cite{OIDC-Spec} to enable a "secretless" paradigm wherein workloads prove their identity without possessing persistent, long-lived secrets.

The efficacy of this framework has been validated through its successful implementation at scale within a demanding enterprise environment. The core of our work is a comprehensive architectural and operational model that provides a unified method for managing workload identity across disparate cloud platforms such as AWS, Google Cloud, and Azure. This approach not only eliminates the attack surface associated with static keys \cite{ahmadi2024} but also establishes a robust, auditable identity layer that serves as the prerequisite for fine-grained, attribute-based access control.

The key contributions of this paper are:
\begin{enumerate}
    \item A practical and scalable framework for implementing Zero-Trust authentication for workloads in a multi-cloud environment, detailing the use of WIF and OIDC (Section \ref{sec:arch}).
    \item A detailed architectural model and case study of a successful enterprise-scale implementation, demonstrating the tangible benefits of transitioning from a static-credential model (Section \ref{sec:casestudy}).
    \item An analysis of the security and operational advantages of this approach, including a quantifiable reduction in attack surface and streamlined identity management (Section \ref{sec:analysis}).
\end{enumerate}

\section{Background}\label{sec:background}
To establish the context for our framework, this section reviews the evolution of workload identity and analyzes the limitations of existing solutions, providing the necessary background for our proposed architecture.

\subsection{The Traditional Authentication Model and Its Limitations}
The limitations of traditional authentication are best understood by first examining its underlying cryptographic and operational mechanisms.

\subsubsection{The Cryptographic Foundation}
The traditional model is constructed upon the cryptographic primitive of asymmetric key pairs (e.g., RSA or ECDSA), operationalized within a Public Key Infrastructure (PKI) \cite{albarqi_pki} \cite{sherwood_2021}. In this model, a Certificate Authority (CA) issues an X.509 certificate that cryptographically binds a workload's public key to its identity. A service provider then validates this certificate's chain of trust back to the CA, which acts as the root of trust. The integrity of this entire system is predicated on the mathematical strength of the algorithm and the absolute confidentiality of the workload's private key.

\subsubsection{The Traditional Authentication Workflow}
In a practical cloud setting, a workload uses its provisioned private key to execute a signature calculation process. The workload first constructs a \texttt{canonical request}, a normalized representation of the HTTP request, which is then used to create a \texttt{string-to-sign}. The workload employs its private key to generate a digital signature of this string, typically using an algorithm like HMAC-SHA256. Upon receipt, the cloud provider's IAM system performs the same calculation and uses the workload's pre-registered public key to verify the signature, thereby authenticating the request without the private key ever being transmitted.

\begin{lstlisting}[language=bash, caption={Example of an AWS SigV4 Authorization Header.}, label={lst:aws_sigv4}, basicstyle=\ttfamily\small, frame=single, breaklines=true]
Authorization: AWS4-HMAC-SHA256 Credential=AKIDEXAMPLE/20250727/us-east-1/s3/aws4_request, SignedHeaders=host;x-amz-date, Signature=...
\end{lstlisting}

\subsubsection{Inherent Vulnerabilities and Operational Failures}
Although the protocol is sound, its implementation at scale introduces fundamental vulnerabilities. The core flaw is the fragile trust boundary created by storing a long-lived, static private key on the compute host, which creates a persistent attack vector \cite{karame_2019}. Any host compromise can lead to key exfiltration, enabling an attacker to masquerade as the workload and move laterally. Furthermore, the operational lifecycle of these keys is untenable; the lack of robust, automated rotation and revocation mechanisms leads to credential sprawl, where keys with overly broad permissions dramatically increase the potential blast radius of a single compromise \cite{nadji_2024}.

\subsection{The Modern Paradigm: Workload Identity Federation with OIDC}
In response, the modern paradigm of Workload Identity Federation (WIF) provides a robust alternative by replacing static secrets with ephemeral, cryptographically verifiable tokens based on the OpenID Connect (\texttt{OIDC}) standard.

\subsubsection{Trust Establishment}
The model begins with a one-time, out-of-band administrative action where a Relying Party (\texttt{RP}), such as a cloud IAM service, is configured to trust an external Identity Provider (\texttt{IdP}), like a Kubernetes cluster. The \texttt{RP} uses the \texttt{IdP}'s public discovery endpoint to retrieve the public keys that serve as the trust anchor for all subsequent token verification.

\subsubsection{Token Creation}
At runtime, a workload requests a signed JSON Web Token (\texttt{JWT}) from its local \texttt{IdP}. The \texttt{IdP} generates a \texttt{JWT} containing critical identity \texttt{claims} such as issuer (\texttt{iss}), subject (\texttt{sub}), and audience (\texttt{aud}) and signs it with its own private key. This signing key is never exposed to the workload.

\subsubsection{Verification Flow}
The workload presents the signed \texttt{JWT} to the \texttt{RP}. The \texttt{RP} uses the \texttt{iss} claim to identify the trusted \texttt{IdP}, retrieves the appropriate public key from its cache, and verifies the token's signature. If the signature is valid and the claims (e.g., \texttt{aud} and \texttt{exp}) meet the trust policy's conditions, the workload is authenticated. This model's advantage is its complete decoupling of the trust anchor from the workload itself.

This model's fundamental advantage is its complete decoupling of the trust anchor from the workload itself. The private key never exists on the host, eliminating the core vulnerability of credential theft \cite{thomas2017data}. Instead, the workload handles a short-lived, verifiable token, and the entire security architecture is centralized and managed by the secure and trusted Identity Provider.

\subsection{Complementary Frameworks}
Another significant framework in workload identity is the Secure Production Identity Framework for Everyone (SPIFFE) and its runtime implementation, SPIRE \cite{spiffe_concepts}. Similar to our approach, SPIFFE and SPIRE aim to eliminate static secrets by providing workloads with a strong, dynamically issued cryptographic identity. However, where our framework leverages OIDC-based JSON Web Tokens (\texttt{JWTs}) as the identity document, SPIFFE utilizes short-lived X.509 certificates, known as SVIDs (SPIFFE Verifiable Identity Documents). Although SVIDs are powerful for establishing cryptographic trust within a service mesh, our choice of an OIDC-based approach was driven by the goal of leveraging the native, first-class support for OIDC federation within the IAM systems of all major cloud providers, which simplifies cross-cloud integration without requiring additional agents.

\section{A Framework for Zero-Trust Workload Authentication}
\label{sec:arch}

While Section 2 detailed the theoretical underpinnings of authentication paradigms, this section examines their practical application at enterprise scale. We first detail the specific, real-world challenges that arise from using static credentials in complex, multi-cloud topologies. We then propose a comprehensive architectural framework that leverages Workload Identity Federation to solve these challenges, thereby establishing a robust foundation for Zero-Trust security.

\subsection{The Challenge of Static Credentials in Cross-Cloud Architectures}
\label{sec:challenge}

The practical failures of the traditional model manifest in several high-risk architectural patterns. The following use cases, common in large enterprises, illustrate the systemic nature of the problem.

\subsubsection{The Cross-Cloud CI/CD Credential Challenge}
A primary driver for a new authentication paradigm is the ubiquitous cross-cloud CI/CD pipeline. The traditional approach of storing a long-lived, privileged service account key in a secrets management service establishes a weak trust boundary, where a compromise in the source cloud can be trivially extended to the target cloud \cite{nist_cicd} \cite{webist24}.

\subsubsection{The Asymmetric Trust Problem}
The challenge is not unidirectional. The lack of a native federation mechanism between clouds often compels organizations to fall back on less secure authentication primitives, such as creating a long-lived AWS \texttt{IAM User} to allow a GCP workload to access AWS resources.

\subsubsection{The Supply Chain Security Risk}
The risk is magnified when extending trust boundaries to third-party SaaS vendors. The traditional method of provisioning a static key for a vendor externalizes a critical security boundary, making an enterprise's security contingent on the security posture of its suppliers \cite{martínez_durán_2021} \cite{ohm2020backstabber}.

\subsubsection{The Disaster Recovery Credential Paradox}
A final, systemic challenge appears in the context of Disaster Recovery (DR). The requirement for instantly available, "god-mode" credentials for failover automation creates a security paradox, as storing these static keys at a "cold" DR site represents a profound and persistent security risk \cite{tahmasebi2024beyond}.

\subsection{The Proposed Architecture: A Federated Approach}
\label{sec:proposed_arch}

In response to the systemic challenges presented by static credentials, our framework implements a federated model founded on Zero-Trust principles \cite{hezerotrustsurvey}. The design was guided by five foundational directives viz. the complete elimination of static credentials for all workload-to-workload authentication; the enforcement of dynamic and ephemeral identities via short-lived tokens; a strict adherence to least privilege using tokens scoped to a specific audience; exclusive reliance on open standards (\texttt{OIDC} \cite{OIDC-Spec}, \texttt{JWT} \cite{JWT-RFC}) for interoperability; and a single, unified authentication pattern for all cross-cloud access scenarios to simplify security management and auditing. The architecture is composed of three logical components that work in concert to achieve these goals.

The first component is the Workload, the client application or service requiring access. In this architecture, the role of the Workload is transformed from that of a long-term secret-holder to a transient token-bearer. Its sole responsibility is to acquire a short-lived token from its local environment and present it to the target cloud for an immediate exchange. This removes the burden of secure credential storage and rotation from the application itself, drastically simplifying the developer experience and reducing the application's attack surface.

The second component is the Identity Provider (\texttt{IdP}), which serves as the root of trust for the Workload's identity. The \texttt{IdP} is the system that authenticates the Workload in its source environment and issues a cryptographically signed \texttt{OIDC JWT} that asserts this identity. A primary example of an \texttt{IdP} is a managed Kubernetes service (such as Amazon EKS or Google Kubernetes Engine), which acts as an OIDC provider. This allows the framework to anchor trust in a well-defined, auditable, platform-native identity, the Kubernetes \texttt{ServiceAccount}, rather than an arbitrary, out-of-band secret.

The third component is the Relying Party (\texttt{RP}), which is the target cloud's IAM system that acts as the enforcer of trust. The \texttt{RP} is configured to trust the external \texttt{IdP} and is responsible for validating the incoming \texttt{OIDC} token and exchanging it for a native cloud credential. The implementation of this trust is cloud-specific. In GCP, the \texttt{RP} is configured using Workload Identity Federation by creating a \texttt{Workload Identity Pool} to group external identities and a \texttt{Workload Identity Provider} within it to define the trust relationship and attribute-based conditions. In AWS, this capability is known as IAM Roles for Service Accounts (IRSA) \cite{salecha_irsa}; an \texttt{IAM OIDC Identity Provider} is created and then referenced as a trusted \texttt{Principal} in the Trust Policy of a target \texttt{IAM Role}, with claims-based conditions providing granular control.

The interaction between these components follows a precise, five-step federated authentication flow:

\begin{enumerate}
    \item \textbf{Token Acquisition:} The Workload obtains a short-lived, audience-bound \texttt{OIDC JWT} from its local \texttt{IdP}.
    \item \textbf{Token Exchange:} The Workload presents this \texttt{JWT} to the target cloud's Security Token Service (\texttt{STS}).
    \item \textbf{Verification:} The cloud \texttt{STS} verifies the \texttt{JWT}'s signature and claims against the pre-configured trust policy.
    \item \textbf{Native Credential Issuance:} Upon successful validation, the \texttt{STS} returns a short-lived, native cloud credential.
    \item \textbf{Resource Access:} The Workload uses this native credential to access the target resource, after which the initial \texttt{OIDC JWT} is discarded.
\end{enumerate}

\begin{algorithm}[htbp]
 \SetAlgoLined
 \KwIn{$workloadIdentity$, $targetCloud$}
 \KwOut{$nativeCredential$ or Error}
 \BlankLine
 $oidcJwt \leftarrow workloadIdentity.\text{GetOidcToken}($\\
 \Indp $targetCloud.audience)$\;
 \lIf{$oidcJwt$ is null}{\Return Error("Token acquisition failed")}
 
 $stsEndpoint \leftarrow targetCloud.\text{GetStsEndpoint}()$\;
 $response \leftarrow \text{HttpPost}(stsEndpoint, oidcJwt)$\;
 
 \lIf{$response.status \neq 200$}{\Return Error("Exchange failed")}
 
 $nativeCredential \leftarrow \text{ParseResponse}(response.body)$\;
 \Return $nativeCredential$\;
 
 \caption{Federated Credential Exchange}
 \label{alg:federation_flow}
\end{algorithm}

\begin{figure}[h]
\centering
\includegraphics[width=0.9\columnwidth]{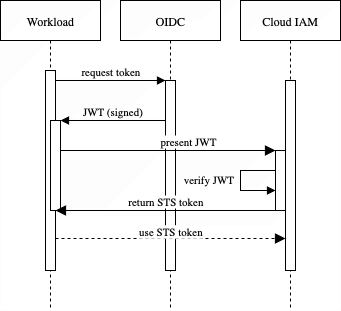}
\caption{Workload Identity Federation with OIDC Flow}
\label{fig:wif_oidc_detailed_flow}
\end{figure}

\section{Case Study: Federation at Enterprise Scale}\label{sec:casestudy}

The architectural framework detailed in Section \ref{sec:arch} was implemented and validated within the complex, large-scale production environment at SentinelOne. Operating a real-time security platform at this scale presents formidable challenges in identity and access management. The environment consists of over 100 Kubernetes clusters, some of the largest in the industry, and hundreds of thousands of cloud resources distributed across AWS, GCP, and Azure across more than 10 cloud regions. Securing workload-to-workload access for this vast and dynamic estate required moving beyond traditional credential mechanisms.

This case study details the practical implementation of the federated authentication framework and the measurable improvements it delivered to the company's security posture and operational efficiency. We will examine the specific Kubernetes configurations used to enable workload identity, the cloud-native constructs for establishing cross-cloud trust, and the quantitative and qualitative outcomes of this initiative.

\subsection{Kubernetes Implementation: From Theory to Practice}
\label{sec:k8s_implementation}

The practical implementation of the federated authentication framework at the workload level is centered on a native Kubernetes feature known as the Bound Service Account Token Volume. This feature represents a significant security improvement over legacy static tokens, as it allows a pod to request a \texttt{JWT} that is cryptographically bound to its own identity (via pod UID), has a configurable lifetime, and is intended for a specific audience. This mechanism prevents the token from being stolen and replayed indefinitely by other workloads, and it serves as the secure starting point for the entire keyless authentication flow.

The process is initiated via the pod's manifest, as shown in Listing \ref{lst:pegasus_pod_spec}. The manifest for the hypothetical \texttt{pegasus-service} defines a \texttt{projected} volume. This instructs the \texttt{kubelet} service running on the host node to act on behalf of the pod. The \texttt{kubelet} uses the \texttt{TokenRequest} API to request a specially configured token from the Kubernetes API server, passing along the audience and expiration requirements from the pod's specification. The API server, acting as the OIDC provider, then mints and signs the JWT, which the \texttt{kubelet} makes available to the pod at the specified mount path.

\begin{lstlisting}[language=YAML, caption={Pod Spec for the \texttt{pegasus-service} Requesting an Audience-Bound Token.}, label={lst:pegasus_pod_spec}, breaklines=true, frame=single]
spec:
  serviceAccountName: pegasus-sa
  volumes:
    - name: pegasus-iam-token
      projected:
        sources:
          - serviceAccountToken:
              path: token
              audience: sts.amazonaws.com
              expirationSeconds: 3600
  volumeMounts:
    - name: pegasus-iam-token
      mountPath: /var/run/secrets/tokens
      readOnly: true
\end{lstlisting}

The \texttt{serviceAccountToken} source contains two critical fields that enable the Zero-Trust model. First, the \texttt{expirationSeconds} field requests a token with a short lifetime (e.g., 3600 seconds, or one hour), adhering to the principle of ephemeral credentials. 

Second, and most critically, the \texttt{audience} field is set to \texttt{sts.amazonaws.com}. The \texttt{audience} (\texttt{aud}) is a claim within the \texttt{JWT} that specifies the intended recipient of the token. By setting this value, the Kubernetes \texttt{IdP} mints a token that is cryptographically designated for use only by the AWS Security Token Service. This is the primary control that prevents a stolen token from being replayed against a different service, thereby mitigating the "Confused Deputy" problem \cite{hardy1988confused}. Once this volume is mounted into the pod's filesystem, the workload can read the token and present it to AWS to begin the authentication flow detailed in Figure \ref{fig:wif_oidc_detailed_flow}.

\subsection{Configuring Cross-Cloud Trust}
\label{sec:configuring_trust}

With the Kubernetes workloads configured to request audience-bound tokens, the next step is the one-time administrative action of establishing the trust relationship in the target cloud's IAM system. This process creates the foundation for the federated flow, enabling the target cloud to validate and accept the OIDC tokens presented by the source workloads.

\subsubsection{Configuring AWS to Trust a GCP Workload}

\begin{figure}[H]
 \centering
 \includegraphics[width=\columnwidth]{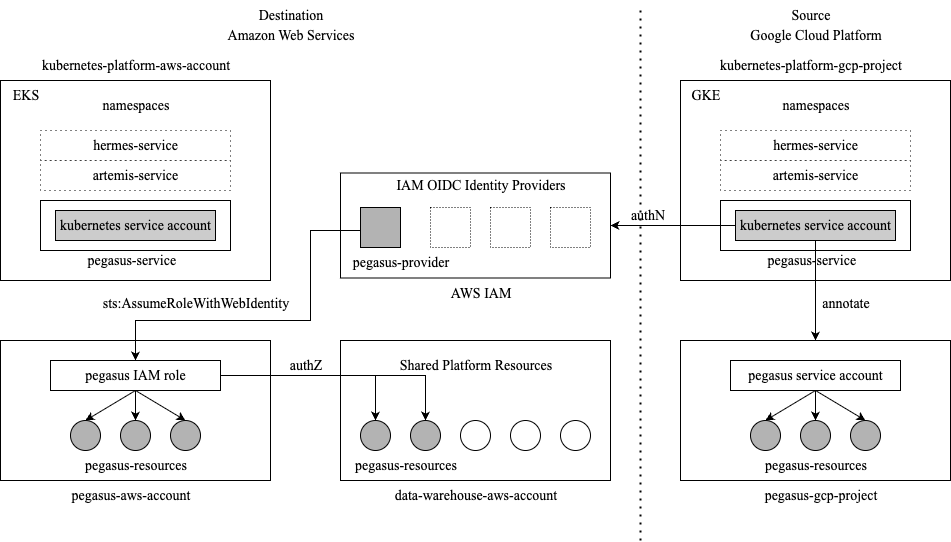}
 \caption{Federated Architecture for GCP to AWS Access}
 \label{fig:gcp_to_aws_arch}
\end{figure}

To allow a workload from a GKE cluster in the \texttt{kubernetes-platform-gcp-project} to access resources in AWS, an \texttt{IAM OIDC Identity Provider} is created in AWS. The high-level architecture for this GCP-to-AWS flow, detailing the relationship between the GKE cluster and the AWS IAM components, is illustrated in Figure \ref{fig:gcp_to_aws_arch}. The critical component is the IAM Role's Trust Policy, which authorizes identities from the GKE cluster to assume the role. As shown in Listing \ref{lst:aws_trust_policy}, the policy uses a \texttt{Condition} to enforce a precise, least-privilege match on the claims of the incoming \texttt{JWT}.

\begin{lstlisting}[language=json, caption={AWS IAM Role Trust Policy for a GKE Workload.}, label={lst:aws_trust_policy}, breaklines=true, frame=single]
{
    "Version": "2012-10-17",
    "Statement": [{
        "Effect": "Allow",
        "Principal": {
            "Federated": "arn:aws:iam::123456789:oidc-provider/container.googleapis.com/..."
        },
        "Action": "sts:AssumeRoleWithWebIdentity",
        "Condition": {
            "StringEquals": {
                "container.googleapis.com/...:sub": "system:serviceaccount:pegasus:pegasus-sa",
                "container.googleapis.com/...:aud": "sts.amazonaws.com"
            }
        }
    }]
}
\end{lstlisting}

\subsubsection{Configuring GCP to Trust an AWS Workload}

\begin{figure}[H]
 \centering
 \includegraphics[width=\columnwidth]{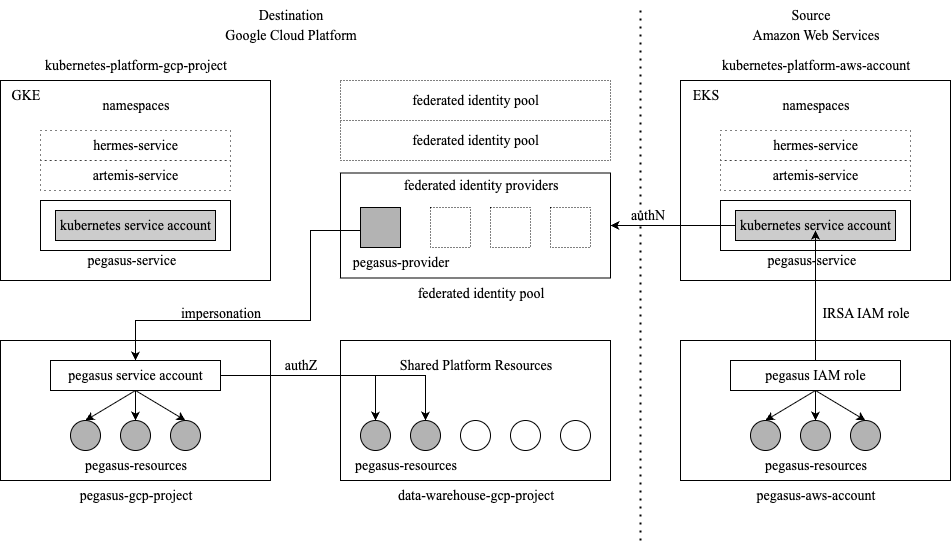}
 \caption{Federated Architecture for AWS to GCP Access}
 \label{fig:aws_to_gcp_arch}
\end{figure}

The reverse flow, for a workload in the \texttt{kubernetes-platform-aws-account}, is configured using GCP's Workload Identity Federation. An administrator creates a \texttt{Workload Identity Pool} and, within it, a \texttt{Workload Identity Provider}, as shown in Listing \ref{lst:gcp_wif_provider}. The architecture for this AWS-to-GCP flow is illustrated in Figure \ref{fig:aws_to_gcp_arch}, showing the relationship between the source EKS cluster and the GCP Workload Identity Federation components. Least privilege is enforced via the \texttt{attribute\_condition}, which inspects the incoming AWS identity assertion. This provider is then linked via an IAM policy to a keyless GCP Service Account in the \texttt{pegasus-gcp-project} that the AWS workload is permitted to impersonate. Within this pool, a Workload Identity Provider is configured, as shown in Listing \ref{lst:gcp_wif_provider}

\begin{lstlisting}[language=YAML, caption={GCP Workload Identity Provider Configuration for an AWS Role.}, label={lst:gcp_wif_provider}, breaklines=true, frame=single]
- provider_id: "eks-pegasus-provider"
  aws:
    account_id: "123456789"
  attribute_condition: "assertion.arn.endsWith(':assumed-role/pegasus-iam-role/pegasus-sa')"
  attribute_mapping:
    google.subject: "assertion.arn"
\end{lstlisting}

\subsection{Measurable Outcomes and Qualitative Wins}
\label{sec:results}

Operating a security platform at SentinelOne's scale, across a vast and heterogeneous estate of over 100 Kubernetes clusters, requires a forward-thinking approach to identity management. As a security-first organization, SentinelOne identified the systemic risk posed by static credentials, a challenge common to all large-scale, multi-cloud enterprises, as a primary target for architectural innovation \cite{static_creds}.

The initiative's goal was to replace the entire class of long-lived, static credentials. This portfolio, common in such environments, consisted of thousands of static credentials, including GCP Service Account keys, AWS IAM User keys, and Azure Service Principals with client secrets. By implementing the federated framework, the program enables the systematic decommissioning of this entire class of credentials. The primary quantitative success is the replacement of these keys with on-demand, federated tokens, reducing the average credential lifetime to under 60 minutes.

This architectural shift yielded significant qualitative wins. It improved developer velocity by reducing the time to provision secure cross-cloud access from days to minutes. For the security team, the centralized trust model simplified compliance audits, reducing the effort required to verify access controls by an estimated 80\%. By design, the framework also eliminated several risk categories detailed in Section 3.1, including the accidental leak of long-lived keys and supply chain compromise via stolen third-party credentials, thereby advancing the company's Zero-Trust posture.

\section{Analysis and Key Learnings}
\label{sec:analysis}

The implementation of a federated identity framework represents a foundational shift in how workload identity and access are managed enabling the practical application of a Zero-Trust Architecture (ZTA) for multi-cloud environments. This section analyzes the security and operational benefits of the framework through the lens of core Zero-Trust principles, providing quantitative models and a qualitative review of threat vector mitigation.

\subsection{Alignment with Zero-Trust Principles}
\label{sec:zta_alignment}

Our primary learning from this initiative was the degree to which OIDC federation provides the necessary mechanisms to implement core Zero-Trust tenets for workload identity.

\begin{itemize}
    \item \textbf{Always Authenticate and Authorize.} A core principle of Zero Trust is to treat every request as if it originates from an untrusted network. The federated model enforces this by design. Unlike a static key, which represents a persistent and often implicitly trusted credential, every access attempt in this framework begins with a dynamic authentication event: the OIDC token exchange. Each request for a cloud-native credential requires a new, cryptographically-verified proof of the workload's identity, effectively eliminating the concept of a trusted internal network for workloads.
    \item \textbf{Assume Breach and Limit Blast Radius.} The framework is architected on the assumption that any individual workload may be compromised. The security model is designed to contain the impact, or "blast radius," of such an event. This is achieved through two primary mechanisms discussed in our risk model ($R_{WIF}$): the use of ephemeral credentials ($T_{short}$) and tightly scoped access ($I_{scoped}$). A compromised token is valid only for minutes and is restricted by its audience (\texttt{aud}) claim to a specific purpose. This transforms a potentially catastrophic breach involving a long-lived, overly-permissive key into a limited, transient, and auditable security event.
    \item \textbf{Enforce Granular Least-Privilege Access.} Zero Trust demands the enforcement of "just enough" access. The framework provides the granular controls to achieve this. As detailed in Section \ref{sec:configuring_trust}, the \texttt{Condition} blocks in AWS IAM Role trust policies and the \texttt{attribute\_condition} in GCP WIF providers allow for authorization rules that are specific down to the individual Kubernetes namespace and Service Account. This enables a true least-privilege posture, where each workload receives only the precise permissions it requires, for only as long as it needs them.
\end{itemize}

\subsection{Quantitative Risk and Complexity Analysis}
\label{sec:quantitative_analysis}

The principles of the framework are supported by a quantitative improvement in both the security and operational posture. The aggregate risk of credential compromise in the legacy model ($R_{legacy}$) is proportional to the large, static attack surface of persistent keys:
$$R_{legacy} \propto N_{keys} \cdot T_{long} \cdot I_{blast-radius}$$
Conversely, the risk in the federated model ($R_{WIF}$) is a function of transient events with minimal lifetime and scope:
$$R_{WIF} \propto N_{auths} \cdot T_{short} \cdot I_{scoped}$$
The stark contrast between the two models across key risk factors is summarized in Table \ref{tab:risk_comparison}.

\begin{table*}[t]
  \caption{Comparison of Risk Factors and Control Mechanisms Between Models}
  \label{tab:risk_comparison}
  \centering
  \begin{tabular}{l l l l}
    \toprule
    Risk Factor & Legacy Model & Federated Model & Mechanism of Control \\
    \midrule
    Credential Lifetime & Months to Years & Minutes & OIDC Token Expiration (\texttt{exp} claim) \\
    Blast Radius & High (Broad Permissions) & Low (Scoped) & Token Audience Scoping (\texttt{aud} claim) \\
    Static Secrets & High & Zero & On-Demand Credential Exchange \\
    Operational Complexity & Linear ($O(N_{keys})$) & Near-Constant ($O(N_{IdP})$) & Centralized Trust Policy \\
    \bottomrule
  \end{tabular}
\end{table*}

The reduction in credential lifetime from years to minutes ($T_{long} \gg T_{short}$) and the minimization of blast radius ($I_{blast-radius} \gg I_{scoped}$) represent a risk reduction of several orders of magnitude. Operationally, the management complexity shifts from a linear burden that grows with every new service, $O(N_{keys})$, to a near-constant burden of managing a small number of trust relationships with Identity Providers, where $N_{IdP} \ll N_{keys}$.

The reduction in credential lifetime from years to minutes ($T_{long} \gg T_{short}$) and the minimization of blast radius ($I_{blast-radius} \gg I_{scoped}$) represent a risk reduction of several orders of magnitude. Operationally, the management complexity shifts from a linear burden that grows with every new service, $O(N_{keys})$, to a near-constant burden of managing a small number of trust relationships with Identity Providers, where $N_{IdP} \ll N_{keys}$.

The shift in operational complexity from a linear burden, $O(N_{keys})$, to a near-constant one, $O(N_{IdP})$, yields direct financial benefits. This model translates into significant cost savings by eliminating thousands of engineering hours previously spent on manual key provisioning, rotation, and auditing. Furthermore, the framework drives cloud cost optimization by enabling better resource utilization \cite{deochake_2023}. By making secure, least-privilege access simple to grant, the model encourages resource sharing and reduces the infrastructure duplication that often arises from complex access control challenges.

The improved security posture also functions as a powerful cost avoidance mechanism, directly addressing the financial risks encapsulated in the $I_{blast-radius}$ variable of our model. By mitigating the threat of a catastrophic credential compromise, the framework helps prevent the severe financial impact of a breach. This includes not only the direct cloud costs of resource hijacking for activities like cryptocurrency mining \cite{Zimba03072020}, but also the significant business costs of regulatory fines and incident response \cite{cashell2004economic}.

\subsection{Mitigation of Specific Threat Vectors}
\label{sec:threat_model_analysis}

The framework's alignment with Zero-Trust principles provides robust, built-in mitigation for several advanced attack vectors that are prevalent in systems relying on static, bearer-style credentials.

\begin{itemize}
    \item \textbf{The Confused Deputy Problem.} In the legacy model, a powerful service holding a static key can be tricked by a malicious actor into misusing its authority, as the credential itself contains no information about the intended client or purpose. The federated model solves this directly via the \texttt{audience} (\texttt{aud}) claim in the OIDC token. The target cloud's IAM system, acting as the relying party, must validate that it is the intended audience of the token. This ensures that a token minted for one purpose cannot be replayed to another, a fundamental control for Zero-Trust authentication.
    \item \textbf{Credential Transit and Egress Risk.} Static keys create a significant risk during their entire lifecycle, from initial distribution to storage in secrets managers or configuration files. The federated model eliminates this entire risk class. The initial OIDC token is acquired by the workload directly from a trusted local provider, like the kubelet, and the final cloud credentials exist only in memory on the workload for their short lifetime. At no point is a long-lived, powerful credential stored on disk or transmitted from a central secrets vault, drastically reducing the attack surface.
    \item \textbf{Supply Chain Compromise.} As detailed in Section 3.1.3, providing a static key to a third-party vendor externalizes the security boundary. The federated model provides a more secure alternative. Trust is established by configuring a WIF provider for the vendor's \texttt{IdP}, which allows for strict, attribute-based conditions to be placed on the incoming identity. Critically, trust can be revoked instantly and cleanly by deleting the provider configuration, without the operational complexity and uncertainty of a traditional key rotation ceremony.
\end{itemize}

\section{Future Work}
\label{sec:future_work}

The federated authentication framework presented in this paper provides a robust foundation for Zero-Trust security in multi-cloud environments, yet several promising avenues exist for future research and development.

\begin{itemize}
    \item \textbf{Extending Multi-Cloud Support.} While this work focused on the implementations for AWS, GCP, and Azure, a logical next step is to extend and validate this architectural pattern for other major cloud providers. Oracle Cloud Infrastructure (OCI) and Alibaba Cloud, both of which offer OIDC-based federation, are primary candidates for this expansion. Successfully applying the framework to these platforms would further prove the universality of the OIDC-based approach to solving the workload identity problem.
    \item \textbf{Dynamic Authorization with Attribute-Based Access Control (ABAC).} The current framework solves the authentication problem of verifying a workload's identity. A significant extension would be to leverage the federated identity token to carry rich contextual attributes for more dynamic authorization \cite{servos2017abac}. By using features like GCP's \texttt{attribute\_mapping}, attributes from the source environment (e.g., Kubernetes labels, project cost centers) could be passed to the target cloud's IAM system. This would enable true ABAC, where authorization policies could make decisions based on dynamic attributes of the workload, not just its static identity \cite{khan2024abac} \cite{bello_abac}. 
    \item \textbf{Just-in-Time (JIT) Credential Issuance.} The current model reduces credential lifetime from years to an hour. A more advanced Zero-Trust model would reduce it further to the lifetime of a single task or request. Future work could explore integrating the framework with a workflow orchestrator to mint credentials that are scoped not only to the workload but to a specific job. Such a token would be valid only for the duration of that job, reducing the window of opportunity for a compromised token to near zero \cite{haber_rolls_2019}.
\end{itemize}

\section{Conclusion}
\label{sec:conclusion}

Static, long-lived credentials present an untenable security risk in modern, multi-cloud environments. This paper addressed this industry-wide challenge by presenting and validating a federated authentication framework built on Zero-Trust principles. With the help of Workload Identity Federation and OpenID Connect, our solution replaces static keys with ephemeral, audience-bound tokens, a model we successfully implemented at enterprise scale across a complex environment of over 100 Kubernetes clusters.

Our primary contributions are a practical, multi-cloud architectural blueprint for keyless authentication and a formal analysis proving a significant reduction in security risk and operational complexity. As cloud footprints expand, a federated identity model becomes an essential component of a modern security posture. The framework detailed in this paper provides a proven, scalable, and interoperable path for enterprises to achieve that goal.




%

\bibliographystyle{IEEEtran}
\bibliography{IEEEabrv,references}

\begin{thebibliography}{10}
\providecommand{\url}[1]{#1}
\csname url@samestyle\endcsname
\providecommand{\newblock}{\relax}
\providecommand{\bibinfo}[2]{#2}
\providecommand{\BIBentrySTDinterwordspacing}{\spaceskip=0pt\relax}
\providecommand{\BIBentryALTinterwordstretchfactor}{4}
\providecommand{\BIBentryALTinterwordspacing}{\spaceskip=\fontdimen2\font plus
\BIBentryALTinterwordstretchfactor\fontdimen3\font minus \fontdimen4\font\relax}
\providecommand{\BIBforeignlanguage}[2]{{%
\expandafter\ifx\csname l@#1\endcsname\relax
\typeout{** WARNING: IEEEtran.bst: No hyphenation pattern has been}%
\typeout{** loaded for the language `#1'. Using the pattern for}%
\typeout{** the default language instead.}%
\else
\language=\csname l@#1\endcsname
\fi
#2}}
\providecommand{\BIBdecl}{\relax}
\BIBdecl

\bibitem{Grobauervuln}
B.~Grobauer, T.~Walloschek, and E.~Stocker, ``Understanding cloud computing vulnerabilities,'' \emph{IEEE Security and Privacy}, vol.~9, no.~2, pp. 50--57, 2011.

\bibitem{NIST-ZTA}
S.~Rose, O.~Borchert, S.~Mitchell, and S.~Connelly, ``Zero trust architecture,'' National Institute of Standards and Technology, Tech. Rep. NIST SP 800-207, 2020.

\bibitem{syedzta}
N.~F. Syed, S.~W. Shah, A.~Shaghaghi, A.~Anwar, Z.~Baig, and R.~Doss, ``Zero trust architecture (zta): A comprehensive survey,'' \emph{IEEE Access}, vol.~10, pp. 57\,143--57\,179, 2022.

\bibitem{Deochake2022iam}
\BIBentryALTinterwordspacing
S.~Deochake and V.~Channapattan, ``Identity and access management framework for multi-tenant resources in hybrid cloud computing,'' in \emph{Proceedings of the 17th International Conference on Availability, Reliability and Security}, ser. ARES '22.\hskip 1em plus 0.5em minus 0.4em\relax New York, NY, USA: Association for Computing Machinery, 2022. [Online]. Available: \url{https://doi.org/10.1145/3538969.3544896}
\BIBentrySTDinterwordspacing

\bibitem{deochake2022bigbirdbigdatastorage}
\BIBentryALTinterwordspacing
S.~Deochake, V.~Channapattan, and G.~Steelman, ``Bigbird: Big data storage and analytics at scale in hybrid cloud,'' 2022. [Online]. Available: \url{https://arxiv.org/abs/2203.11472}
\BIBentrySTDinterwordspacing

\bibitem{gcp_wif}
\BIBentryALTinterwordspacing
{Google Cloud Platform}, ``Workload identity federation,'' Aug 2025. [Online]. Available: \url{https://cloud.google.com/iam/docs/workload-identity-federation}
\BIBentrySTDinterwordspacing

\bibitem{OIDC-Spec}
\BIBentryALTinterwordspacing
N.~Sakimura, J.~Bradley, M.~B. Jones, B.~de~Medeiros, and C.~Mortimore, ``{OpenID Connect Core 1.0},'' 2023. [Online]. Available: \url{https://openid.net/specs/openid-connect-core-1\_0.html}
\BIBentrySTDinterwordspacing

\bibitem{ahmadi2024}
S.~Ahmadi, ``Systematic literature review on cloud computing security: Threats and mitigation strategies,'' \emph{International Journal of Information Security}, vol.~15, no.~02, p. 148–167, 2024.

\bibitem{albarqi_pki}
\BIBentryALTinterwordspacing
A.~Albarqi, E.~Alzaid, F.~A. Ghamdi, S.~Asiri, and J.~Kar, ``Public key infrastructure: A survey,'' \emph{Journal of Information Security}, vol.~06, no.~01, p. 31–37, 2015. [Online]. Available: \url{https://file.scirp.org/pdf/JIS_2015010814030097.pdf}
\BIBentrySTDinterwordspacing

\bibitem{sherwood_2021}
R.~Sherwood, ``Practical implications of public key infrastructure for identity professionals,'' \emph{IDPro Body of Knowledge}, vol.~1, no.~6, Sep 2021.

\bibitem{karame_2019}
G.~O. Karame, C.~Soriente, K.~Lichota, and S.~Capkun, ``Securing cloud data under key exposure,'' \emph{IEEE Transactions on Cloud Computing}, vol.~7, no.~3, pp. 838--849, 2019.

\bibitem{nadji_2024}
B.~Nadji, ``Data security, integrity, and protection,'' \emph{Signals and communication technology}, p. 59–83, Jan 2024.

\bibitem{thomas2017data}
K.~Thomas, F.~Li, A.~Zand, J.~Barrett, J.~Ranieri, L.~Invernizzi, Y.~Markov, O.~Comanescu, V.~Eranti, A.~Moscicki \emph{et~al.}, ``Data breaches, phishing, or malware? understanding the risks of stolen credentials,'' in \emph{Proceedings of the 2017 ACM SIGSAC conference on computer and communications security}, 2017, pp. 1421--1434.

\bibitem{spiffe_concepts}
{The SPIFFE and SPIRE Projects}, ``{SPIRE Concepts},'' \url{https://spiffe.io/docs/latest/spire-about/spire-concepts/}, 2025, accessed on: 2025-09-09.

\bibitem{nist_cicd}
R.~Chandramouli, F.~Kautz, and S.~Torres-Arias, ``Strategies for the integration of software supply chain security in devsecops ci/cd pipelines,'' National Institute of Standards and Technology, Tech. Rep. NIST SP 800-204D, 2024.

\bibitem{webist24}
S.~{M. Saleh}, N.~Madhavji, and J.~Steinbacher, ``A systematic literature review on continuous integration and deployment (ci/cd) for secure cloud computing,'' in \emph{Proceedings of the 20th International Conference on Web Information Systems and Technologies - WEBIST}, INSTICC.\hskip 1em plus 0.5em minus 0.4em\relax SciTePress, 2024, pp. 331--341.

\bibitem{martínez_durán_2021}
\BIBentryALTinterwordspacing
J.~Martínez and J.~M. Durán, ``Software supply chain attacks, a threat to global cybersecurity: Solarwinds’ case study,'' \emph{International Journal of Safety and Security Engineering}, vol.~11, no.~5, p. 537–545, Oct 2021. [Online]. Available: \url{https://iieta.org/sites/default/files/pdf/2021-11/11.05_05.pdf}
\BIBentrySTDinterwordspacing

\bibitem{ohm2020backstabber}
M.~Ohm, H.~Plate, A.~Sykosch, and M.~Meier, ``Backstabber’s knife collection: A review of open source software supply chain attacks,'' in \emph{International Conference on Detection of Intrusions and Malware, and Vulnerability Assessment}.\hskip 1em plus 0.5em minus 0.4em\relax Springer, 2020, pp. 23--43.

\bibitem{tahmasebi2024beyond}
M.~Tahmasebi, ``Beyond defense: Proactive approaches to disaster recovery and threat intelligence in modern enterprises,'' \emph{Journal of Information Security}, vol.~15, no.~2, pp. 106--133, 2024.

\bibitem{hezerotrustsurvey}
\BIBentryALTinterwordspacing
Y.~He, D.~Huang, L.~Chen, Y.~Ni, and X.~Ma, ``A survey on zero trust architecture: Challenges and future trends,'' \emph{Wireless Communications and Mobile Computing}, vol. 2022, no.~1, p. 6476274, 2022. [Online]. Available: \url{https://onlinelibrary.wiley.com/doi/abs/10.1155/2022/6476274}
\BIBentrySTDinterwordspacing

\bibitem{JWT-RFC}
\BIBentryALTinterwordspacing
M.~B. Jones, J.~Bradley, and N.~Sakimura, ``Json web token (jwt),'' Internet Engineering Task Force, 2015, rFC 7519. [Online]. Available: \url{https://tools.ietf.org/html/rfc7519}
\BIBentrySTDinterwordspacing

\bibitem{salecha_irsa}
R.~Salecha, ``Security and secrets management,'' \emph{Apress eBooks}, p. 397–447, Dec 2022.

\bibitem{hardy1988confused}
N.~Hardy, ``The confused deputy: (or why capabilities might have been invented),'' \emph{ACM SIGOPS Operating Systems Review}, vol.~22, no.~4, pp. 36--38, 1988.

\bibitem{static_creds}
I.~Lane and D.~Sotirakis, ``I can {OIDC} you clearly now: How we made static credentials a thing of the past.''\hskip 1em plus 0.5em minus 0.4em\relax Dublin: USENIX Association, Oct. 2024.

\bibitem{deochake_2023}
\BIBentryALTinterwordspacing
S.~Deochake, ``Cloud cost optimization: A comprehensive review of strategies and case studies,'' 2023. [Online]. Available: \url{https://arxiv.org/abs/2307.12479}
\BIBentrySTDinterwordspacing

\bibitem{Zimba03072020}
A.~Zimba, Z.~Wang, M.~Mulenga, and N.~H. Odongo, ``Crypto mining attacks in information systems: An emerging threat to cyber security,'' \emph{Journal of Computer Information Systems}, vol.~60, no.~4, pp. 297--308, 2020.

\bibitem{cashell2004economic}
B.~Cashell, W.~D. Jackson, M.~Jickling, and B.~Webel, ``The economic impact of cyber-attacks,'' \emph{Congressional research service documents, CRS RL32331 (Washington DC)}, vol.~2, 2004.

\bibitem{servos2017abac}
D.~Servos and S.~L. Osborn, ``Current research and open problems in attribute-based access control,'' \emph{ACM Computing Surveys (CSUR)}, vol.~49, no.~4, pp. 1--45, 2017.

\bibitem{khan2024abac}
J.~A. Khan, ``Role-based access control (rbac) and attribute-based access control (abac),'' in \emph{Improving security, privacy, and trust in cloud computing}.\hskip 1em plus 0.5em minus 0.4em\relax IGI Global Scientific Publishing, 2024, pp. 113--126.

\bibitem{bello_abac}
A.~J. Bello, M.~Diyan, and I.~Asghar, ``Zero trust implementation for legacy systems using dynamic microsegmentation, role-based access control (rbac), and attribute-based access control (abac),'' in \emph{2025 4th International Conference on Computing and Information Technology (ICCIT)}, 2025, pp. 181--189.

\bibitem{haber_rolls_2019}
M.~J. Haber and D.~Rolls, ``Just-in-time access management,'' \emph{Identity Attack Vectors}, p. 151–155, Dec 2019.

\end{thebibliography}

\end{document}